# Deep Music Retrieval for Fine-Grained Videos by Exploiting Cross-Modal-Encoded Voice-Overs

Tingtian Li[1], Zixun Sun[1], Haoruo Zhang[1], Jin Li[1], Ziming Wu[1], Hui Zhan[1], Yipeng Yu[1], Hengcan Shi[2]
[1]Interactive Entertainment Group, Tencent
[2]University of Electronic Science and Technology of China
tingtian.li@outlook.com; {zixunsun, haoruozhang}@tencent.com; j.lixjtu@gmail.com;
zwual@connect.ust.hk; {huizhan, ianyu}@tencent.com; shihengcan@gmail.com

## ABSTRACT

Recently, the witness of the rapidly growing popularity of short videos on different Internet platforms has intensified the need for a background music (BGM) retrieval system. However, existing video-music retrieval methods only based on the visual modality cannot show promising performance regarding videos with fine-grained virtual contents. In this paper, we also investigate the widely added voice-overs in short videos and propose a novel framework to retrieve BGM for fine-grained short videos. In our framework, we use the self-attention (SA) and the cross-modal attention (CMA) modules to explore the intra- and the inter-relationships of different modalities respectively. For balancing the modalities, we dynamically assign different weights to the modal features via a fusion gate. For paring the query and the BGM embeddings, we introduce a triplet pseudo-label loss to constrain the semantics of the modal embeddings. As there are no existing virtual-content video-BGM retrieval datasets, we build and release two virtual-content video datasets HoK400 and CFM400. Experimental results show that our method achieves superior performance and outperforms other state-of-the-art methods with large margins.

## 1 INTRODUCTION

Video-music retrieval is an emerging topic and recently obtains increasing attention due to the global popularity of short videos. Many short video Apps like Instagram, WeSee, and TikTok, provide the function of adding BGM to the edited video before the video uploading. A proper video-music retrieval system can retrieve reasonable BGM for ordinary users to enhance the video phenomena and lower the video production barrier. In the meanwhile, with the flourishment of the virtual entertainments, it is reported that there are 2.7 billion video gamers [1] and the views of the top 5 most viewed games on YouTube have achieved 485 billion in 2020 [2]. Videos related to virtual content have gained substantially increasing popularity, especially during the Covid-19 pandemic. As the virtual-content videos are visually only composed of some specific 3D models and most user-generated videos are only associated with a limited number of popular games, the videos of a game or some close environments are visually very similar even the semantics of the videos are very different. Pioneering methods retrieve BGM for videos via matching their emotion tags [3, 4]. Then, [5] and [6] use the conventional classifiers to match the hand-crafted modal features. Recently, [7, 8] develop two-stream frameworks of deep neural networks (DNNs) to retrieve BGM for videos in an end-to-end way. However, these methods that only use the visual modality show limited performance for the virtual-content videos due to the similar and fine-grained visual frames.

Nowadays the voice-over is a pervasively used skill in short video production. The voice-over in a short video can be intentionally emotional and is used to grab the attention of the audience. It significantly impacts the semantics of the final edited video. For example, the voice-over can show an exciting tune for a coming highlight and turn to be odd at a funny moment. The voice-over is designed according to the author intention and usually written in the script with the way to clip the video. It is an impactive factor to choose the final BGM. In this paper, we propose a novel framework named MRCMV that retrieves BGM for the fine-grained virtual-content videos when voice-overs are ready. The voice-over is a critical hint to retrieve BGM, when the visual contents are very fine-grained. The framework is shown in Fig. 1. For the query branch, we first use the SlowFast [9] and the VGGish [10] networks to extract the video and the voice-over features respectively. Then, we use the SA and the CMA modules to exploit the intra- and the inter-modal relationships. Besides, we introduce a fusion gate to dynamically balance the modal contributions. For the BGM branch, we only use the SA module to exploit the intra-modal relationship. Thus, the BGM modality is isolated from the querying modalities and all the BGM embeddings can be obtained before the query comes. As there are no existing virtual-content video-music datasets, we release two virtual-content short video datasets HoK400 and CFM 400 for evaluation. To summarize, the contributions of this paper are as follows:

1) To our knowledge, we are the first to use both the visual and the voice-over modalities to conduct the video-music retrieval task. We also propose using the SA and the CMA modules to explore their intra- and inter-modal relationships respectively. The following fusion gate can learn adaptive weights to balance the modal embeddings.

2) To match different modal embeddings, we propose a triplet pseudo-label loss to constrain the semantic similarities of the paired embeddings. Experimental results show that it can improve the retrieving performance rather than only using the triplet loss.

3) As there are no existing virtual-content video-music retrieval datasets, we build and release two such datasets HoK400 and CFM400. The videos are collected from public short video platforms and made by professional teams. We hope the datasets can promote the progress of video-music retrieval research in the information retrieval community.

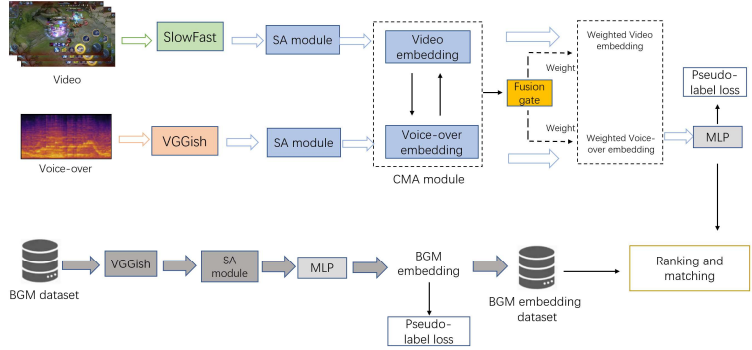

Figure 1: The framework of the proposed algorithm MRCMV. The SA and the CMA modules explore the intra- and the inter-modal relationships. The fusion gate assigns adaptive weights to balance the contributions of different modalities. The triple-pseudo-loss constrains the semantic similarities of paired embeddings.

## 2 RELATED WORK

In terms of video-music retrieval, conventional works bridge the visual and auditory features through video emotion tags [3, 4]. However, human-defined subjective emotion tags easily introduce accumulated biases to the two modal sides. [5] and [6] try to use the expectation-maximization (EM) algorithm [11-13] and the support vector machine (SVM) [14, 15] respectively to match different modalities. But the hand-crafted features cannot give the exact representations w.r.t. various modal semantics. [16] learns the visual features using the convolutional neural networks (CNNs). But they still use hand-crafted features to extract the BGM features.

---
*The datasets are at https://github.com/pixel-333/MRCMV

Recently, inspired by the image-text retrieval frameworks, which have been intensively studied in [17-20], [7, 8] develop the two-stream frameworks for the video-music retrieval task. [7] first trains a two-stream CNN on an emotion tag dataset and then fine-tunes the networks for modality matching. However, the emotion tags are very coarse and subjective. They are not suitable for fine-grained videos. [8] adds a soft intra-modal structure to keep the embedding internal-specific. Nonetheless, it is still difficult to constrain the videos with fine-grained contents.

## 3 THE PROPOSED METHOD

Fig. 1 shows the proposed BGM retrieval framework contains two branches, which generate the query joint embedding and the candidate BGM embedding respectively. In this section, we will elaborate on the details of the embedding generations.

### 3.1 The Joint Embedding Generation

For the joint embedding generation branch, we first use the SlowFast [9] and the VGGish [10] networks to extract the video feature $F = fc([F^{fast}, F^{slow}])$ and the voice-over feature $S = [s_0, s_1, ..., s_N]$, where $F^{fast} = [f_1^{fast}, f_2^{fast}, ..., f_N^{fast}]$ and $F^{slow} = [f_1^{slow}, f_2^{slow}, ..., f_N^{slow}]$. $f_i^{fast}$ and $f_i^{slow}$ are the $i$th frame features extracted by the fast and the slow paths of the SlowFast network respectively; $s_i$ is the $i$th frame feature extracted by the VGGish network; $N$ is the number of frames; $fc$ is a fully-connected layer, $[.,.]$ is the concatenation operation. As $F$ and $S$ are in different domains, we need to transfer them into the same space.

### 3.1.1 Modal Relationship Exploiting

The features within a single modality are not temporarily independent. A visual or a voice-over frame can be related to other frames in the video sequence. Besides, the visuality and the voice-over also have inter-modal relationships. The emotions and tunes of the voice-over can change along with different visual contents. Inspired by the transformer that is originally used to explore the contextual representation [21], we extend the multi-head attention mechanism and use it to explore the intra- and the inter-relationships of the visual and the voice-over modalities as follows

$$O_i(I_1, I_2) = softmax\left(\frac{I_1 W_i^Q (I_2 W_i^K)^T}{\sqrt{d_k}}\right) I_2 W_i^V, \quad (1)$$

$$\chi(I_1, I_2) = I + [O_0(I_1, I_2), \ldots, O_{N_h}(I_1, I_2)] W^O, \quad (2)$$

where $\chi(I_1, I_2)$ is a general module to explore the representation between features $I_1 \in R^{N_1 \times d_{I_1}}$ and $I_2 \in R^{N_2 \times d_{I_2}}$; $W_i^Q \in R^{d_{I_1} \times d_k}$, $W_i^K \in R^{d_{I_2} \times d_k}$ and $W_i^V \in R^{d_{I_2} \times d_v}$ are learnable matrices; $W^O \in R^{N_h d_v \times d_k}$ is the learnable matrix to encode the multi-head concatenated result; $N_h$ is the number of heads. Here, $d_{I_1} = d_{I_2}$ and $d_{I_1} = N_h d_v$ for feature alignment. When $I_1 = I_2$, $\chi$ degrades to the SA module. The self-attended visual and voice-over features can be obtained by $\tilde{F} = SA(F) = \chi(F, F)$ and $\tilde{S} = SA(S) = \chi(S, S)$ respectively. When $I_1 \neq I_2$, $\chi$ becomes the CMA module. It learns the inter-modal representation and updates the modal features as $\dot{F} = CMA(\tilde{F}, \tilde{S}) = \chi(\tilde{F}, \tilde{S})$ and $\dot{S} = CMA(\tilde{S}, \tilde{F}) = \chi(\tilde{S}, \tilde{F})$.

### 3.1.2 A Fusion Gate to Balance the Modalities

The intuitive ways to fuse the features are concatenation and summarization. However, the direct operations treat the modal features equally and cannot give an adaptive fusion. Therefore, we propose a fusion gate to dynamically balance the contributions of $\dot{F}$ and $\dot{S}$ as follows

$$g(\dot{F}, \dot{S}) = \sigma\left(([\dot{F}, \dot{S}]) W^g + b^g\right), \quad (3)$$

$$H_J(\dot{F}, \dot{S}) = g(\dot{F}, \dot{S})_{\{:d_{\dot{F}}\}} \circ \dot{F} + g(\dot{F}, \dot{S})_{\{d_{I_{\dot{S}}}+1:\}} \circ \dot{S}, \quad (4)$$

where $W^g \in R^{(d_{\dot{F}}+d_{\dot{S}}) \times (d_{\dot{F}}+d_{\dot{S}})}$, $b^g \in R^{1 \times (d_{\dot{F}}+d_{\dot{S}})}$ and $\sigma$ is the ReLU activation function; $\circ$ means the element-wise multiplication; $g(\dot{F}, \dot{S})_{\{:d_{\dot{F}}\}}$ is the cropped elements along the depth axis from 0 to $d_{\dot{F}}$ and $g(\dot{F}, \dot{S})_{\{d_{\dot{F}}+1:\}}$ is the cropped elements from $d_{\dot{F}} + 1$ to the end. We can find the gate is trained to assign adaptive weights to each dimension of $\dot{F}$ and $\dot{S}$. Then we feed the fused feature to the next fully connected layer after reshaping the depth of $H_J(\dot{F}, \dot{S})$ to be 1. We also normalize the result and finally obtain the $1 \times d_E$ joint embedding $E_J$.

### 3.2 The BGM Embedding Generation

For decoupling the two branches of the retrieval system, the BGM embedding branch only has one modality. Consequently, we only use the SA module to explore the intra-relationship of the BGM features and obtain the self-attended feature $\tilde{M} = SA(M)$, where $M$ is the BGM feature which is obtained as the same as $S$. Like the joint embedding, we reshape $\tilde{M}$ and feed it to a fully connected layer. Finally, we normalize it and obtain the $1 \times d_E$ BGM embedding $E_M$.

### 3.3 The Triplet Pseudo-Label Loss

For matching the embeddings, the triplet loss is often used to shrink the distance of each paired embeddings. In this paper, we also consider using the supervised manner to constrain the embedding semantics. It has been proved that the network intermediate features for classification tasks can represent semantics [22-25]. Thus, we assign the same labels to the paired embeddings and assign different labels to negative embeddings. The total loss function $\mathcal{L}$ becomes

$$\mathcal{L} = \mathcal{L}_1 + \lambda \mathcal{L}_2, \quad (5)$$

$$\mathcal{L}_1 = \sum_i \sum_k max\left(\left\|E_J^i - E_M^i\right\|_2^2 - \left\|E_J^i - E_M^k\right\|_2^2 + \alpha, 0\right), \quad (6)$$

$$\mathcal{L}_2 = \sum_i \sum_k \left(-log\left(\frac{e^{E_J^i W^J(z_i)}}{\sum_c^N e^{E_J^i W^J(z_c)}}\right) - log\left(\frac{e^{E_M^i W^M(z_i)}}{\sum_c^N e^{E_M^i W^M(z_c)}}\right) - log\left(\frac{e^{E_M^k W^M(z_k)}}{\sum_c^N e^{E_M^k W^M(z_c)}}\right)\right), \quad (7)$$

where $\mathcal{L}_1$ is the triplet loss and $\mathcal{L}_2$ is the proposed triplet pseudo-label loss; $\lambda = 0.1$ is a balancing constant; $E_J^i$ and $E_M^i$ mean the $i$th paired joint embedding and the $i$th paired BGM embedding respectively; $z_i$ is the pseudo label for the $i$th paired embeddings; $\alpha$ is the margin in the triplet loss; $W^J$ and $W^M$ are two learnable matrices to classify the pseudo labels for the joint embedding and the BGM embedding respectively. Notice $\mathcal{L}_2$ forces the paired embeddings to point to the same label and share similar semantics. All the network parameters including the feature extraction parts are trained together with the loss function $\mathcal{L}$.

## 4 EXPERIMENTS

### 4.1 Datasets

As there are no existing video-music datasets that focus on virtual contents, we build and release two virtual-content video datasets HoK400 and CFM400 for experiments. The HoK400 and the CFM400 contain 427 and 401 online short videos related to the globally popular video games 'King of Honors' (alias 'Arena of Valor') and 'CrossFire Mobile' respectively. All the videos with both voice-overs and BGM are collected from short video platforms and elaborately made by professional short video production teams. The BGM are pure music without singing voices for retaining clear voice-overs. Thus, we can use the vocal separator Spleeter [26] to separate BGM and voice-overs from the audio track. Then, we train the networks in a self-supervised manner that matches the query of videos with voice-overs back to their original BGM. We split the datasets and train the networks using 265 and 270 videos from the HoK400 and the CFM400 datasets respectively. The test datasets contain the rest of 162 and 131 videos.

Table 1: The evaluation results on different datasets

| Method | HoK400 | | | | | CFM400 | | | | |
|---|---|---|---|---|---|---|---|---|---|---|
| | R@1 | R@5 | R@10 | R@25 | VGGish Distance | R@1 | R@5 | R@10 | R@25 | VGGish Distance |
| CMMR | 1.2 | 3.7 | 9.3 | 24.1 | 32.7 | 3.1 | 8.4 | 14.5 | 34.4 | 31.6 |
| CBVMR | 1.2 | 6.8 | 13.0 | 29.0 | 32.1 | 0.8 | 8.4 | 13.0 | 29.0 | 33.6 |
| MRCMV w/o the SA modules | 32.7 | 59.2 | 67.9 | 84.0 | 22.0 | 28.2 | 57.3 | 73.3 | 91.6 | 23.7 |
| MRCMV w/o the CMA modules | 23.8 | 54.9 | 67.1 | 88.4 | 24.0 | 20.6 | 48.9 | 61.1 | 81.7 | 26.4 |
| MRCMV w/o the fusion gate | 21.0 | 49.3 | 70.3 | 85.8 | 25.4 | 24.4 | 52.7 | 67.9 | 90.1 | 24.9 |
| MRCMV w/o the triplet pseudo-label loss | 21.0 | 42.6 | 53.1 | 80.2 | 25.6 | 29.8 | 59.5 | 75.6 | 93.1 | 23.8 |
| MRCMV | **37.7** | **62.3** | **79.6** | **93.8** | **19.4** | **35.8** | **75.5** | **83.2** | **94.7** | **21.2** |

Table 2: The preference results of the subjective evaluation

| | CMMR | CBVMR | MRCMV |
|---|---|---|---|
| HoK400 | 28.4% | 16.0% | **55.6%** |
| CFM400 | 20.8% | 25.6% | **53.6%** |

Table 3: The preference results of the added songs

| | Search under tags | Neutral | MRCMV with HoK400 |
|---|---|---|---|
| Preferred | 19.2% | 38.4% | **42.3%** |

## 4.2 Implementation Details

We use the stochastic gradient descent (SGD) to train the networks with the learning rate $1 \times 10^{-6}$ and the momentum 0.9. The embedding dimension $d_E$ is 512. The margin $\alpha$ in (6) is constantly equal to 30 at the first 80 epochs and increases by 10 every 10 epochs until the loss converges. We randomly cut the videos in the training dataset into 32 seconds clips for data augmentation. For the test part, we use the fixed beginning 32 seconds clips to evaluate the performance. If the video length is less than 32 seconds, we simply repeat the video until its length is longer than 32 seconds. We treat the clips in a video with distances below 7s intervals that have similar semantics and assign the same pseudo labels to them. We sample 64 and 8 frames from the videos and input them to the fast and the slow paths of the SlowFast network respectively. For the voice-overs and the BGM, we both sample 72 frames from their Mel-spectra before feeding them to the VGGish networks.

## 4.3 Ablation Study and Comparisons

For evaluating the proposed method, we compare the proposed method to the recent methods CMMR [7] and CBVMR [8], and conduct an ablation study to verify the effectiveness of each component. The results are shown in Table 1. Following the multi-modal one-to-one retrieval tasks [8, 20], we use the Recall at K (R@K, K=1, 5, 10, and 25) to measure the percentage of the queries for which the ground truth was ranked among the top K matches. We also use the VGGish intermediate feature that is trained on another irrelevant large-scale audio classification dataset Audioset [27] to measure semantic differences between the top-1 recalled BGM and the ground truth. As the datasets are irrelevant, the parameters of the network for similarity measuring are very different from our audio feature extraction network. Table 1 shows the proposed method with all the components gives the best performance. This verifies the effectiveness of the proposed components to explore the modal representations. It is also noticeable that our method can significantly outperform the competing methods CMMR and CBVMR which only use the visual modality as the query. This shows our method with the voice-over modality can significantly improve the performance. We also conduct subjective evaluations. We ask humans to compare the anonymous results of the test videos with the BGM retrieved by the different methods. Table 2 shows the percentages of each method they preferred, where our method still shows the best performance. We also demonstrate the proposed framework trained by our datasets without the singing voices for the self-supervised training can be used for retrieving songs for videos, although the singing voices in the songs may disturb the clear voice-overs. We use the Spleeter to obtain pure music parts of the songs as the retrieving candidates. The vocal part will be added back if the retrieved BGM originally has the singing voice. As it is difficult to split the voice-over and singing voice from an audio track of a fully edited video, we cooperate with a professional 'King of Honors' short video production team. They produce 26 short videos with voice-overs but without BGM. They would like to try the new BGM retrieval system and compare their originally used random search under selected music style tags that fit the video phenomena. The results of the two groups are anonymous. The candidate pool has 320 songs. Here, the MRCMV is trained by the HoK400 training dataset. Table 3 shows the preference results of the professional team. We can see the MRCMV trained by the dataset without singing voices for the self-supervised training still retrieves better songs even the random search uses the music style tags.

## 5 CONCLUSIONS

In this paper, we propose a novel BGM retrieval framework for fine-grained virtual-content videos. We combine the voice-over with the conventional visual modality to retrieve BGM and use the SA and the CMA modules to explore the intra- and the inter-modal relationships. For balancing the modalities, we use a fusion gate to adaptively assign weights to each dimension of the embeddings. For better pairing the embeddings, we introduce the triplet pseudo-label loss to further constrain their semantics. Together with the framework, we also build and release two fine-grained video datasets HoK400 and CFM400. Experimental results show that our method achieves superior performance and outperforms other state-of-the-art methods.

# REFERENCES

[1] T. Wijman. 2020. The World's 2.7 Billion Gamers Will Spend $159.3 Billion on Games in 2020; The Market Will Surpass $200 Billion by 2023. *Newzoon*.

[2] R. Wyatt. 2020. 2020 is YouTube Gaming's biggest year, ever: 100B watch time hours. *YouTube Officail Blog*.

[3] K.-H. Shin and I.-K. Lee. 2017. Music synchronization with video using emotion similarity. In *Proceedings of The IEEE International Conference on Big Data and Smart Computing*, 47-50.

[4] J. Chao, H. Wang, W. Zhou, W. Zhang, and Y. Yu. 2011. Tunesensor: A semantic-driven music recommendation service for digital photo albums. In *Proceedings of the International Semantic Web Conference*.

[5] E. Brochu, N. De Freitas, and K. Bao. 2003. The sound of an album cover: Probabilistic multimedia and information retrieval. In *Workshop on Artificial Intelligence and Statistics*.

[6] X. Wu, Y. Qiao, X. Wang, and X. Tang. 2016. Bridging music and image via cross-modal ranking analysis. *IEEE Transactions on Multimedia*, vol. 18, no. 7, pp. 1305-1318.

[7] B. Li and A. Kumar. 2019. Query by Video: Cross-modal Music Retrieval. In *Proceedings of the International Society for Music Information Retrieval*, 604-611.

[8] S. Hong, W. Im, and H. S. Yang. 2018. Cbvmr: content-based video-music retrieval using soft intra-modal structure constraint. In *Proceedings of the ACM on International Conference on Multimedia Retrieval*, 353-361.

[9] C. Feichtenhofer, H. Fan, J. Malik, and K. He. 2019. Slowfast networks for video recognition. In *Proceedings of the IEEE International Conference on Computer Vision*, 6202-6211.

[10] S. Hershey, S. Chaudhuri, D.P.W. Elis *et al.*. 2017. CNN architectures for large-scale audio classification. In *Proceedings of the IEEE International Conference on Acoustics, Speech and Signal Processing*, 131-135.

[11] M. A. Figueiredo and R. D. Nowak. 2003. An EM algorithm for wavelet-based image restoration. *IEEE Transactions on Image Processing*, vol. 12, no. 8, 906-916.

[12] A. P. Dempster, N. M. Laird, and D. B. Rubin. 1977. Maximum likelihood from incomplete data via the EM algorithm. *Journal of the Royal Statistical Society: Series B (Methodological)*, vol. 39, no. 1, 1-22.

[13] T. Li, D. P. K. Lun, and T.-W. Shen. 2015. Improved expectation-maximization framework for speech enhancement based on iterative noise estimation. In *Proceedings of the IEEE International Conference on Digital Signal Processing*, 287-291.

[14] T. Evgeniou and M. Pontil, 1999, Support vector machines: Theory and applications. *Advanced Course on Artificial Intelligence*, 249-257.

[15] W. S. Noble. 2006. What is a support vector machine? *Nature biotechnology*, vol. 24, no. 12, 1565-1567.

[16] E. Acar, F. Hopfgartner, and S. Albayrak. 2014. Understanding affective content of music videos through learned representations. In *Proceedings of International Conference on Multimedia Modeling*, 303-314.

[17] Y. Jian, J. Xiao, Y. Cao, A. Khan, and J. Zhu. 2019. Deep Pairwise Ranking with Multi-label Information for Cross-Modal Retrieval. In *Proceedings of IEEE International Conference on Multimedia and Expo*, 1810-1815.

[18] P. Hu, L. Zhen, D. Peng, and P. Liu. 2019. Scalable deep multimodal learning for cross-modal retrieval. In *Proceedings of the International ACM SIGIR Conference on Research and Development in Information Retrieval*, 635-644.

[19] L. Zhen, P. Hu, X. Wang, and D. Peng, 2019, Deep supervised cross-modal retrieval. In *Proceedings of the IEEE Conference on Computer Vision and Pattern Recognition*, 10394-10403.

[20] X. Yang, J. Dong, Y. Cao, X. Wang, M. Wang, and T.-S. Chua. 2020. Tree-Augmented Cross-Modal Encoding for Complex-Query Video Retrieval. In *Proceedings of the International ACM SIGIR Conference on Research and Development in Information Retrieval*, 1339-1348.

[21] A. Vaswani, N. Shazeer, N. Parmar *et al.*. 2017. Attention is all you need. *Advances in neural information processing systems*, 5998-6008.

[22] J. Johnson, A. Alahi, and F.-F. Li. 2016. Perceptual losses for real-time style transfer and super-resolution. In *Proceedings of the European Conference on Computer Vision*, 694-711.

[23] T. Li, Y.-H. Chan, and D. P. K. Lun. 2020. Improved Multiple-Image-Based Reflection Removal Algorithm Using Deep Neural Networks. *IEEE Transactions on Image Processing*, vol. 30, 68-79.

[24] C. Ledig, L. Theis, F. Huszar *et al.*. 2017. Photo-realistic single image super-resolution using a generative adversarial network. In *Proceedings of the IEEE Conference on Computer Vision and Pattern Recognition*, 4681-4690.

[25] T. Li and D. P. K. Lun. 2019. Single-image reflection removal via a two-stage background recovery process. *IEEE Signal Processing Letters*, vol. 26, no. 8, 1237-1241.

[26] R. Hennequin, A. Khlif, F. Voituret, and M. Moussallam. 2020. Spleeter: a fast and efficient music source separation tool with pre-trained models. *Journal of Open Source Software*, vol. 5, no. 50, 2154, 2020.

[27] J. F. Gemmeke, D. P. W. Ellis, D. Freedman *et al.*. 2017. Audio set: An ontology and human-labeled dataset for audio events In *Proceedings of the IEEE International Conference on Acoustics, Speech and Signal Processing*, 776-780.